\documentclass[conference]{IEEEtran}
\usepackage{cite}
\usepackage[cmex10]{amsmath}
\usepackage{amssymb}
\usepackage{graphicx}
\usepackage{epstopdf}
\usepackage{graphics}
\usepackage{pdfpages}
\usepackage{float}
\usepackage{multirow}
\usepackage{epsfig,color,algorithm}
\usepackage{algpseudocode}
\usepackage{cases}
\usepackage{changes}
\usepackage{algorithmicx}

\ifCLASSINFOpdf
\else
\fi
\usepackage[cmex10]{amsmath}
\usepackage{epsfig} 
\ifCLASSOPTIONcompsoc
  \usepackage[caption=false,font=normalsize,labelfont=sf,textfont=sf]{subfig}
\else
  \usepackage[caption=false,font=footnotesize]{subfig}
\fi
\usepackage{fixltx2e}
\usepackage{stfloats}
\begin{document}


\title{Blind Numerology Identification for Mixed Numerologies}

\author{
\IEEEauthorblockN{Ahmad Jaradat\IEEEauthorrefmark{1}, Ebubekir Memi\c{s}o\u{g}lu\IEEEauthorrefmark{1}, and H\"{u}seyin Arslan\IEEEauthorrefmark{1}\IEEEauthorrefmark{2}}
\IEEEauthorblockA{\IEEEauthorrefmark{1}Department of Electrical and Electronics Engineering, Istanbul Medipol University, Istanbul, 34810, Turkey\\
\IEEEauthorblockA{\IEEEauthorrefmark{2}Department of Electrical Engineering, University of South Florida, Tampa, FL, 33620, USA\\
Email: ahmad.jaradat@std.medipol.edu.tr, ebubekir.memisoglu@std.medipol.edu.tr, huseyinarslan@medipol.edu.tr}
}
}


\maketitle
\begin{abstract}
5G New Radio (NR) introduces {new} flexibility that different numerologies can be selected to meet the requirements of a wide variety of services. For this new structure, blind numerology identification can increase system efficiency. Therefore, we propose a blind identification method for mixed numerologies. An autocorrelation method is applied in the time domain by correlating the cyclic prefix (CP) signal of the {candidate} numerology in the received composite signal {for numerology {type} identification. Then, the location of each numerology in the frequency domain is identified by the variance difference in the power spectral density (PSD) of the subbands}, on which different numerologies are occupied. The simulation results are obtained {under} additive white Gaussian noise (AWGN) and frequency-selective {channels}. {The obtained results show that the proposed method has a robust identification accuracy} {and a {satisfactory} BER performance as compared to the {non-blind identification approach in the} conventional mixed-numerology system.}
\end{abstract} 

%
\IEEEpeerreviewmaketitle
\section{Introduction}\label{intsec}
\IEEEPARstart{T}{he} Third Generation Partnership Project (3GPP) adopted orthogonal frequency division multiplexing (OFDM) as the 5G New Radio (NR) waveform for downlink and uplink transmissions. {An OFDM signal is a rectangular symbol composed of orthogonal subcarriers that carry the modulated symbols. 
{The frequency spacing between the orthogonal subcarriers {equals} the reciprocal of the symbol time.} The set of subcarrier spacings (SCSs) and symbol duration {in 5G} {is} {flexible} and referred to as an OFDM numerology. { This mixed-numerology structure has} been studied in \cite{Guan2017,Iwabuchi2017,Zaidi2016CM} and included in the NR standards {of} 3GPP \cite{3GPP2018}.}

The structure of {mixed-numerology offers} attractive flexibility in 5G and beyond networks. As the adjustable parameters of numerology increase, it becomes more critical to find the optimum one \cite{YazarFlexibility2018}. 
Due to this variety in the parametrization of numerologies, distinguishing them by the user equipments (UEs) becomes crucial in the 5G NR.
The UEs can identify numerology using different approaches; for instance, UE can know what to look for, or it can exploit higher-layer signaling (of known numerology) for informing the UE, or it can also employ a blind detection method.

Blind detection algorithms are used to increase spectral efficiency by estimating the signal parameters without  additional signals. 
There are some identification methods proposed to identify OFDM signals \cite{Walter2000,Liedtke2008}. 
These methods target conventional OFDM with single numerology. Therefore, they cannot be directly reflected in the mixed-numerology system. 

The idea of blind numerology identification was presented in \cite{20190190647}. This work explains how to identify different numerologies in time and frequency domains. For this goal, the cyclic prefix (CP) correlation in the time domain and power spectral density (PSD) mask in the frequency domain are exploited. 
With the assumption of the use of blank resources, the numerologies are identified by measuring the size and disposition of their blank resources in the PSD mask.
Therefore, only numerology identification is aimed {at \cite{20190190647}} without identifying its locations in the frequency domain. It is worth noting that this work proposed identification methods without focusing on specific communication systems. For example, the flexible mixed-numerology structure with mixed numerologies has not been investigated. Furthermore, the performance evaluation of this work has not been conducted.

To the best of our knowledge, no study in the literature explains the blind numerology identification for mixed-numerology transmissions. 
{We} propose a {new} method for blind numerology identification, where numerology types and their subband locations can be identified. For this purpose, two receiver algorithms are developed by employing the CP correlation-based and signal variance-based approaches in the time and frequency domains, respectively. The performance of the proposed {method} is examined under additive white Gaussian noise (AWGN) and frequency-selective channels. Our proposed method achieves high {identification accuracy} without any prior information. Also, the BER performance of the proposed method is obtained and compared with the {non-blind identification approach in the conventional} mixed-numerology OFDM system.

{Our main contributions are summarized as follows:}
\begin{itemize}
    \item {We propose a novel method to solve the {blind identification problem} of different numerologies for a mixed numerologies system without prior information.}
    \item {Low-complexity receiver algorithms are jointly proposed for numerology type and location identification based on its characteristics in time and frequency domains. We adopt the CP autocorrelation and subband amplitude variance algorithms in the time and frequency domains, respectively.} 
    \item {The conducted simulations show the efficiency of the proposed method in terms of identification accuracy and BER performances under different environments, including AWGN and multipath fading channels.}
\end{itemize}

The structure of this paper is organized as follows: Section II presents our system model. Section III provides the proposed identification method. Section IV discusses the obtained simulation results. Finally, Section V concludes the paper.

\section{System Model}\label{syssecmod}
The base numerology adopted in Long Term Evolution (LTE) has $\Delta f_0=15$ kHz SCS.
\thinspace In our analysis, we assume numerology with the subcarrier spacing as \cite{3GPP166364}

\begin{equation}
\Delta f_k=2^{k} \Delta f_0,    
\end{equation}
where {$k=\{0, 1, 2, 3, ...\}$ is the scaling factor of a given numerology.}
\thinspace {The inverse fast Fourier transform (IFFT)/FFT size of a scalable OFDM numerology can be written as}
\begin{equation}
N_k=2^{-k} N_0,     
\end{equation}
{where $N_0$ {denotes} IFFT/FFT size of the reference numerology.}
{The symbol duration of the scalable numerology} can be expressed as

\begin{equation} \label{tofdm}
T_{OFDM,k}=T_{DATA,k}+T_{CP,k},     
\end{equation}
{where {data duration} $T_{DATA,k}=1/\Delta f_k$ and {CP duration {varies for each numerology with $T_{DATA,k}$ as}} $T_{CP,k}=\alpha \thinspace T_{DATA,k}$ for $0<\alpha<1$.}
\thinspace{The duration of the OFDM symbol for the reference numerology ($T_{OFDM,0}$) is related to $T_{OFDM,k}$ as}

\begin{equation}
T_{OFDM,k}=2^{-k} \thinspace T_{OFDM,0}.    
\end{equation}
{{The sampling rate $T_{s}$ for all used numerologies is considered the same as} \cite{3GPP166364}}

\begin{equation}
T_{s}=\Delta f_k \thinspace N_k=\Delta f_0 \thinspace N_0.    
\end{equation}

\begin{figure}[t]
\centering
\includegraphics[height=2.5in,width=3.4in]{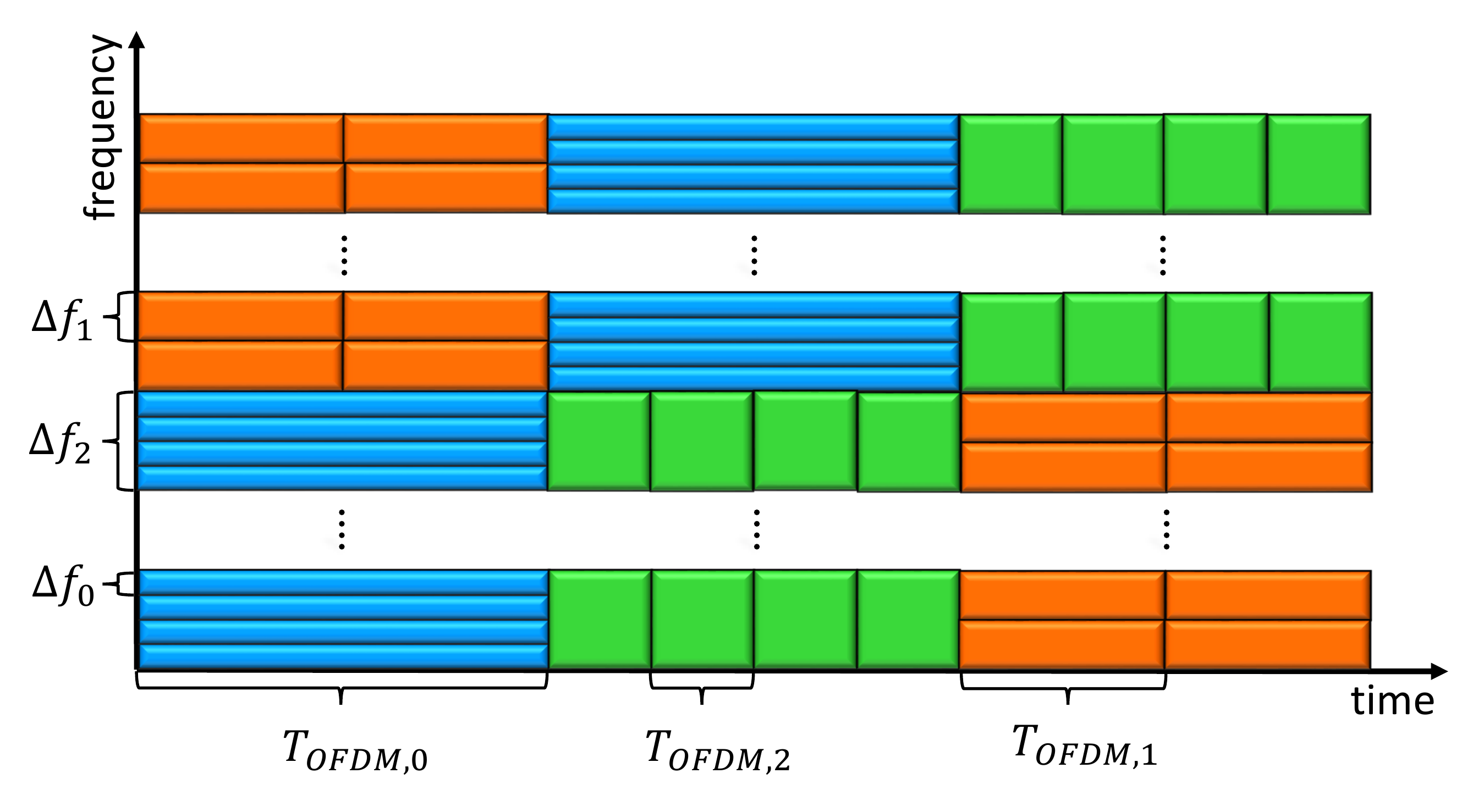}

\caption{Frame structure of mixed-numerology system.}
\label{sysmod}
\end{figure}

The time and frequency domain representations of three different numerologies with $4\Delta f_0 = 2\Delta f_1 =\Delta f_2$ are presented in Fig. \ref{sysmod}. {We assume that} users with different numerologies share the {system bandwidth} equally. Therefore, each subband of different numerologies {has the} same bandwidth \cite{Guan2017}. 

The vector of  modulated symbols for {the} $u$-th {user}, {where $u\in\{1, 2, ..., U\}$}, in the frequency domain can be expressed by

\begin{equation}
    \mathbf{s}_u= [s_u(1)\ s_u(2)\ ... \ s_u(M_k) \ 0 \ ... \ 0]_{1\times N_k}^T,
\end{equation}
where $M_k =2^{-k} M_0$ represents the total number of active subcarriers for the $u$-th user ($M_0$ is the number of activated subcarriers of the reference numerology).

The used numerologies are represented by the set of 
$\chi =  \{ \text{Numerology} \ 1, \text{Numerology} \ 2,...,\text{Numerology} \ U\}$. Here, each user {uses distinct numerology with a specific time slot that} occupies different subbands {of} the whole band.  Then, the time-domain signal is obtained by taking IFFT as 

\begin{equation}
    \mathbf{x}_u = \frac{1}{\sqrt{N_k}}\mathbf{W}_{N_k}^H\mathbf{s}_u,
\end{equation}
where $\mathbf{W}_{N_k}$ represents the inverse discrete Fourier transform (IDFT) matrix with $\mathbf{W}_{N_k}^H \mathbf{W}_{N_k} = N_k \thinspace \mathbf{I}_{N_k}$. Here, $\mathbf{I}_{N_k}$ is an $N_k \times N_k$ identity matrix.

To prevent intersymbol interference, the CP with a length of $N_{CP,k}$ is added to $\mathbf{x}_u$. {The {complete} CP-OFDM signal of the $u$-th user ($\mathbf{x}_{CP,u}$) {is obtained} by concatenating multiple CP added symbols.} {Then, the $u$-th user sends the $\mathbf{x}_{CP,u}$ signal} over a wireless channel {with a channel impulse response represented by $\mathbf{h}_u$}. This channel-impaired received signal is contaminated with AWGN represented by the vector $\mathbf{w}$. Finally, the noise-contaminated {OFDM} signal received from $U$ users can be represented as

\begin{equation}
    \mathbf{y} = \sum_{u=1}^{U} \mathbf{x}_{CP,u} \ast \mathbf{h}_u + \mathbf{w},
\end{equation}
where the received vector $\mathbf{y}$ has a length of $N+N_{CP}$ ($\smash{\displaystyle N=\max_{k}}  \thinspace \{N_k\}$ and $N_{CP}=\alpha N$).

\section{Proposed numerology identification method}
In this section, we discuss the proposed identification algorithms based on the characteristics of numerology transmission. The autocorrelation method is commonly used in the literature for signal detection. We adopted a CP correlation method based on the periodicities brought by the CP signals in different numerologies. This approach is built based on the fact that the maximum point of the time-domain metric corresponds to the starting point of an OFDM symbol. For each numerology, we can independently obtain different timing metrics.

{We assume that the receiver has no prior information about numerology parameters. More specifically, SCS, data/CP durations are not known at the receiver. Therefore, we employ our method with all possible numerology parameters.}
To identify the candidate numerology for the $u$-th user, the received composite {OFDM} signal ($\mathbf{y}$) is correlated with all possible CP signals in $\mathbf{y}$.
The absolute values of the normalized correlation of $\mathbf{y}$ for the $u$-th user at the position $i=n_s \; \mathrm{mod} \; (N_k+N_{CP,k})$  are calculated by (9), where $n_s=0, 1, ..., N+N_{CP}-1$ represents the sample index.

\newcounter{mytempeqncnt}
\begin{figure*}
\centering
\footnotesize
\setcounter{mytempeqncnt}{10}
\setcounter{equation}{8}
\setlength{\arraycolsep}{0.5em}
\begin{eqnarray}
\begin{split}
\tilde{C}_{CP}(k,i)=\frac{\left|\sum_{j=1}^{N_{CP,k}} y\left[n_s+j\right]^{*}y\left[n_s+N_k+j\right]\right|}{\sqrt{\sum_{j=1}^{N_{CP,k}}\left|y\left[n_s+j\right]\right|^{2}}\sqrt{\sum_{j=1}^{N_{CP,k}}\left|y\left[n_s+N_k+j\right]\right|^{2}}}
\end{split}
\label{cpcorr}
\end{eqnarray}
\setlength{\arraycolsep}{5pt}
\setcounter{equation}{9}
\hrulefill
\normalsize
\end{figure*}

We consider finding the first two highest peaks in the first OFDM symbol of each numerology as
\begin{equation} 
\begin{split}
i_{p1}&=\arg \max_{i_1} \left\{\tilde{C}_{CP}(k,i_1)\right\}, \\
i_{p2}&=\arg \max_{i_2} \left\{\tilde{C}_{CP}(k,i_2)\right\}, 
\end{split}
\end{equation}
where $i_1=0,1, ...,(N_k+N_{CP,k})/2-1$ and $i_2=(N_k+N_{CP,k})/2, (N_k+N_{CP,k})/2+1 ...,(N_k+N_{CP,k})-1$. 
{Afterward, the calculated distance between the peaks of the candidate numerology can be found as follows:} 
\begin{equation} \Delta f_{\hat{k}}=|i_{p2}-i_{p1}|. \end{equation}
If $\Delta f_{\hat{k}}$ corresponds to the numerology with $\Delta f_k$ whose order is defined in the set $\chi$, then a correct identification of numerology type as $\Delta f_{\hat{k}}$ is achieved.
The proposed time-domain identification algorithm is summarized in {Algorithm \ref{TDalg}}.

\alglanguage{pseudocode}
\begin{algorithm}[h!]
\small
\caption{The proposed time-domain identification algorithm}
\label{TDalg}
\begin{algorithmic}[1]
\State $ \tilde{C}_{CP}(k,i) \gets $ Correlate the CP signal for the $u$-th user in $\mathbf{y}$.
\State $i_{p1}, i_{p2} \gets$ Evaluate the positions of the first two highest correlation\vspace{1mm} peaks as in (10).
\State $\Delta f_{\hat{k}}=|i_{p2}-i_{p1}| \gets $ Estimate the FFT size of the candidate numerology. 
\If {$\Delta f_{\hat{k}} \in \chi$}
\State Correct identification of numerology type.
\Else
\State Incorrect identification of numerology type.
\EndIf
\Statex
\end{algorithmic}
\vspace{-0.4cm}%
\end{algorithm} 

The frequency-domain offers reliable results about signals in the communication systems.
After identifying the numerology in the time domain using Algorithm 1, we can detect its location by exploiting its characteristics in the frequency domain.
The time-domain samples are transferred to their corresponding frequency-domain ones. 

\thinspace The CP of {the numerology, whose type is identified in the time domain,} {is removed} from $\mathbf{y}$, and the resultant signal given as $\mathbf{y}_{k}$.
\thinspace Afterward, the FFT {operation} with a length of $N_k$ is employed to $\mathbf{y}_k$ to get the subband signal of the $u$-th user in the frequency domain ($\mathbf{y}_{F,u}$). 
{The location of the {candidate} numerology {is detected} by observing the lowest {amplitude} {variation in} the absolute response of $\mathbf{y}_{F,u}$ (i.e. $|\mathbf{y}_{F,u}|$) {for the $U$  subbands}.} 
\;{More specifically, the variation coefficient ($V_u$) {is calculated} for each subband  as}
\begin{equation}
\begin{split}
V_u&=\sigma/\mu\\ &
=\dfrac{\frac{1}{N_k-1} \sum_{u}^{}\left||\mathbf{y}_{F,u}|-\frac{1}{N_k} \sum_{u}^{} |\mathbf{y}_{F,u}|\right|^{2}}{\frac{1}{N_k} \sum_{u}^{} |\mathbf{y}_{F,u}|},
\end{split}
\end{equation}
where $\sigma$ and $\mu$ represent the {sample} variance and mean over a specific band in $|\mathbf{y}_{F,u}|$ with a length of $N_k$. 
The location of the {candidate} numerology {is estimated based on the obtained $V_u$ values as}

\begin{equation} {\hat{u}} = \arg \min_{u} \left\{V_u\right\}. \end{equation}
From (13), the numerology of the $\hat{u}$-th user subband is identified as $\Delta f_k$. Since the FFT operation is employed based on $N_k$, the $\hat{u}$-th user subband with the minimum $V_u$ value identifies the numerology for the ${u}$-th user  subband as  $\Delta f_k^{\hat{u}}$. 
If there is no mismatch between numerologies of the users as $\Delta f_k^{\hat{u}} = \Delta f_k^{u}, \forall u $, then a correct identification of numerology location is achieved. Otherwise, an incorrect identification  occurs, and it introduces extra bit errors besides the error bits in the detection of the conventional symbols in the classical mixed-numerology system. 
\;{The step-by-step procedure of the proposed frequency-domain algorithm can be summarized, as shown in {Algorithm \ref{FDalg}}.}

\vspace{-0.2cm}%
\alglanguage{pseudocode}
\begin{algorithm}[h!]
\small
\caption{The proposed frequency-domain identification algorithm}
\label{FDalg}
\begin{algorithmic}[1]
\State $ \mathbf{y}_{k} \gets $ Remove the CP signal part of the {candidate numerology} from $\mathbf{y}$.
\State $\mathbf{y}_{F,u} \gets $ Employ FFT with length $N_k$ to $\mathbf{y}_{k}$.
\State $|\mathbf{y}_{F,u}| \gets $ Find the amplitude response of $\mathbf{y}_{F,u}$.
\State $V_u \gets $ Find the variation coefficient for each band in $|\mathbf{y}_{F,u}|$.
\State $\hat{u} =\arg \min_{u} \left\{V_u\right\} \gets $ Estimate the location of the {candidate} numerology.
\If {$\Delta f_k^{\hat{u}} = \Delta f_k^{u}, \forall u$}
\State Correct identification of numerology location.
\Else
\State Incorrect identification of numerology location.
\EndIf
\Statex
\end{algorithmic}
\vspace{-0.4cm}%
\end{algorithm} 

To evaluate the robustness of the proposed method, its BER performance is compared with the theoretical results of the conventional OFDM that are formulated for binary phase shift keying (BPSK) transmissions over AWGN and Rayleigh fading channels as \cite{proakis2008digital}
\begin{equation}P_{b}=\frac{1}{2} \operatorname{Q}(\sqrt{2 SNR})\end{equation}
and
\begin{equation}P_{b}=\frac{1}{2}\Bigg(1-\sqrt{\frac{SNR}{SNR+1}}\Bigg),\end{equation}
where $Q(.)$ is the standard Q-function \cite{Chiani2002} and $SNR=E_b/N_{o,T}$ ($E_b$ and $N_{o,T}$ represent the bit energy and the noise variance in the time domain, respectively).

The computational complexity is evaluated per detected numerology for the proposed method and the non-blind identification approach.
The autocorrelation method adopted in Algorithm 1 (step 1) involves $N_{CP,k}$ multiplications and additions for a limited number of lags. 
The maximum operation used in step 2 of Algorithm 1 has a complexity order of $\mathcal{O}(N_k+N_{CP,k})$.
Step 3 in Algorithm 1 requires constant time complexity. The complexity of the conditional operation (step 4 to 8 in Algorithm 1) is based on $U$.
The FFT, variation coefficients calculations, minimum, and conditional operations are performed in Algorithm 2. The complexity order of FFT is $\mathcal{O}(N_k \thinspace \log_2 N_k)$ \cite{cooley1965algorithm}, and $U\times M_k$ additions/subtractions are needed to find the variation coefficients for the whole used band. Minimum and conditional operations have a complexity of the order $\mathcal{O}(U)$. Thus, the overall complexity level of the proposed method is linear-logarithmic.

\section{Simulation Results}
In this section, the simulation results are illustrated to investigate the robustness of the {proposed method}. \thinspace{In the conducted simulations, the received mixed-numerology signals are assumed to be affected by AWGN and multipath Rayleigh fading channels.} The considered multipath channel is characterized by an impulse response of length 9 and normalized power delay profile $\boldsymbol\varrho=[0.8407, 0, 0, 0.1332, 0, 0.0168, 0.0067, 0, 0.0027]$ mW \cite{cho2010mimo}.
For each SNR value, we performed the simulations $10^4$ times to ensure good accuracy.

For the sake of simplicity, we considered two scenarios as Scenario 1 and Scenario 2 in the conducted simulations. The base numerology with 15 kHz SCS is mixed with 30 kHz SCS and 60 kHz SCS numerologies in Scenario 1 and Scenario 2, respectively. The adopted numerology parameters are shown in Table \ref{numparam} \cite{3GPP2018}.
The CP ratio is set as $\alpha=1/16$. 
We assume an uplink transmission of two users using different numerologies with BPSK symbols carried on the activated subcarriers.

\begin{table}[!h]
\begin{center}
\caption{The numerologies parameters in the considered scenarios}
\begin{tabular}{|c|c|c|c|}
\hline
\textbf{Numerology} & NUM-1 & NUM-2 & NUM-3 \\ \hline
$k$ & 0 & 1 & 2 \\ \hline
$\Delta f_k$ & 15 kHz & 30 kHz & 60 kHz \\ \hline
$N_k$ & 4096 & 2048 & 1024 \\ \hline
$N_{CP,k}$ & 256 & 128 & 64 \\ \hline
$M_k$ & 1024 & 512 & 256 \\ \hline
Number of OFDM symbols & 1 & 2 & 4 \\ \hline
\end{tabular}
\label{numparam}
\end{center}
\end{table}

For a clear presentation of Algorithm 1, its performance has shown without considering the channel distortion and noise effects.
Fig. 2 displays the conducted simulations for Scenario 1. Correct identification of the 15 kHz and 30 kHz SCSs candidate numerologies is achieved in the time domain, as shown in Fig. 2. The average distances between the consecutive highest peaks are 4096 and 2048 samples, which correspond to the IFFT/FFT sizes of the 15 kHz and 30 kHz SCSs candidate numerologies, respectively.

\begin{figure}[t]
             \begin{center}
              \subfloat[15 kHz numerology]{\label{CPcorrwithout1:1}\includegraphics[width=45mm]{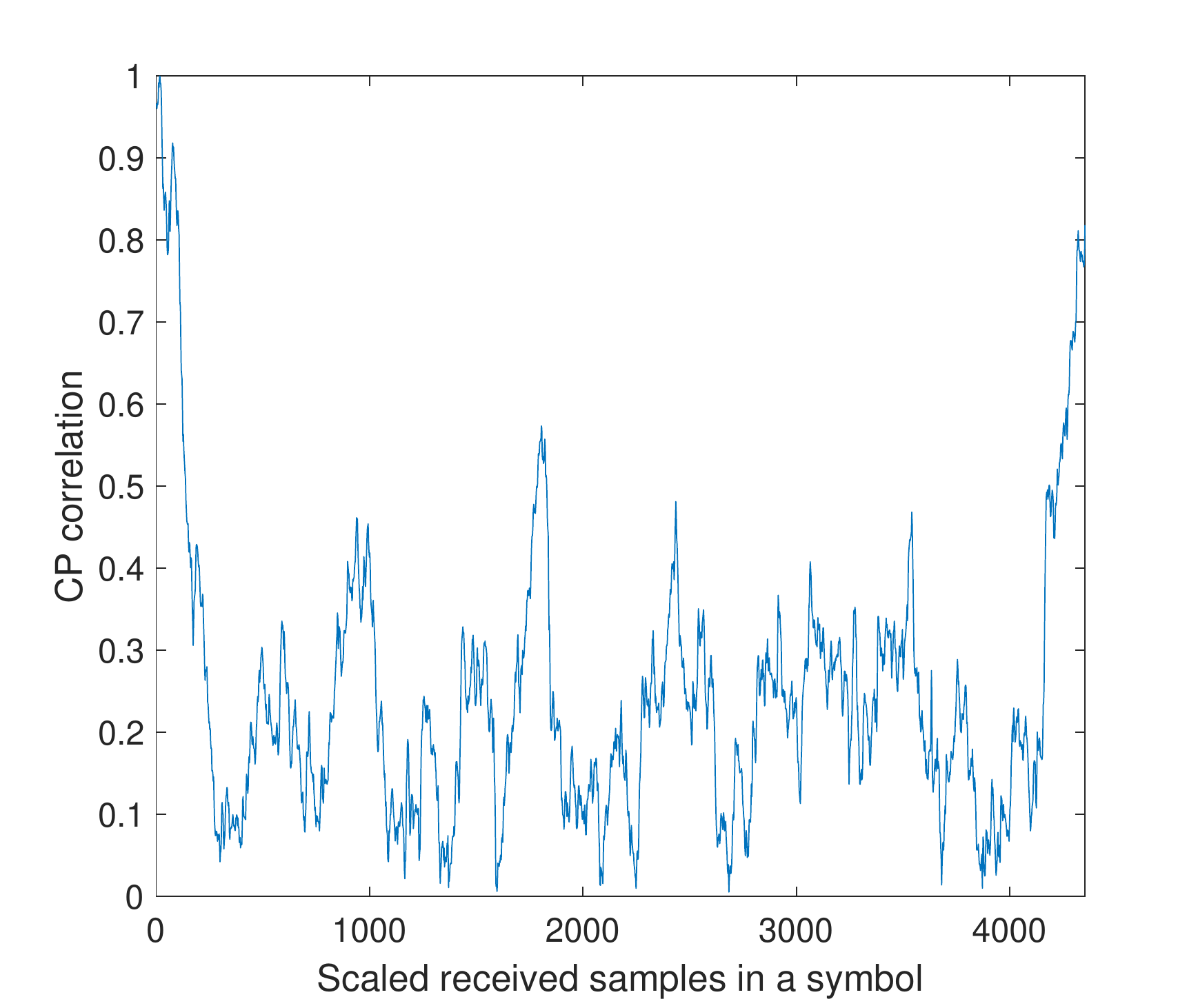}}
               \subfloat[30 kHz numerology]{\label{CPcorrwithout1:2}\includegraphics[width=45mm]{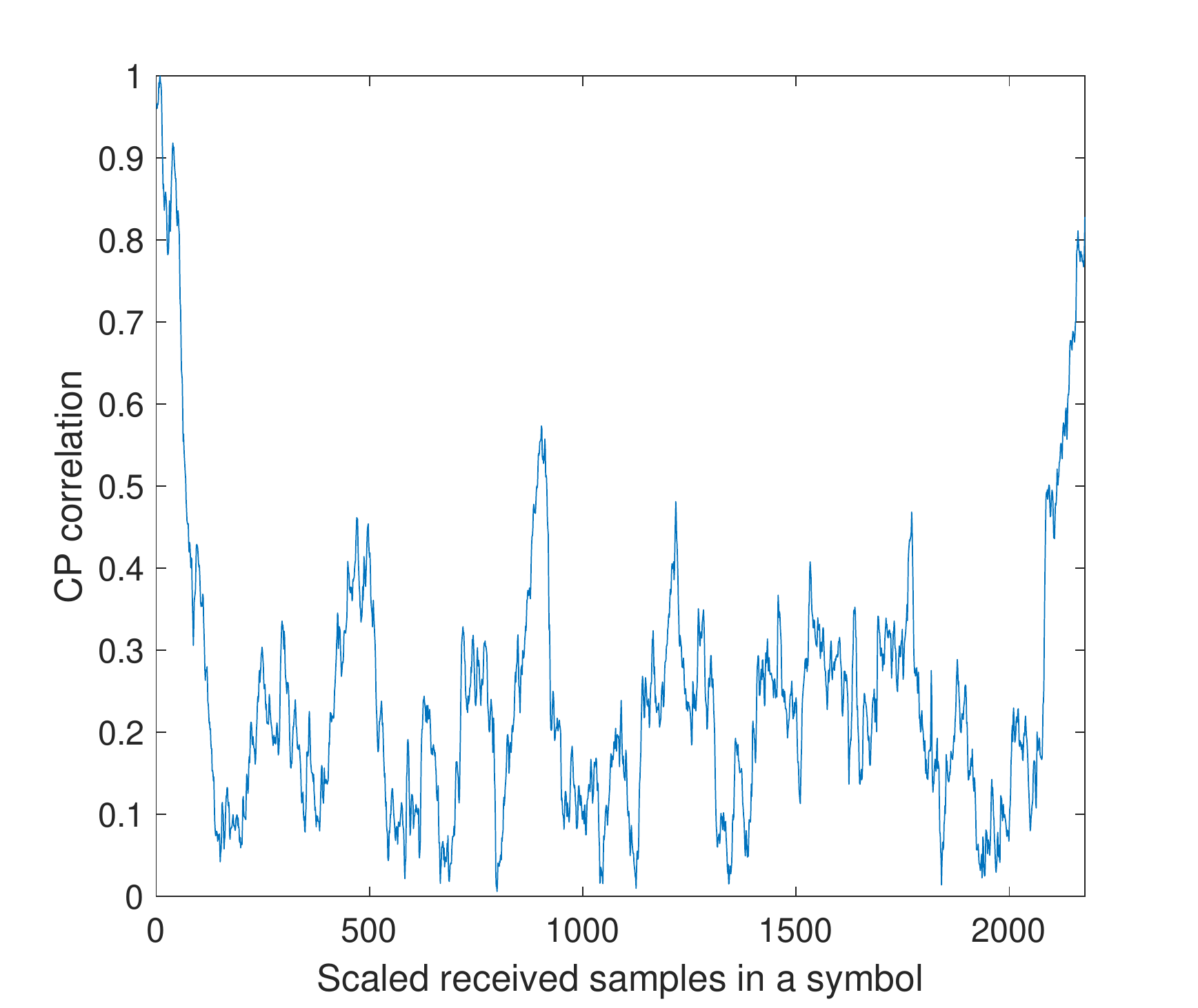}}
                \\
                \end{center}
                 \centering
           \label{CPcorrwithout1}\caption{CP correlation calculations for Scenario 1 without considering noise and channel effects.}
\end{figure}

\begin{figure}[t]
             \begin{center}
              \subfloat[15 kHz numerology]{\label{fftwithout15khz:1}\includegraphics[width=45mm]{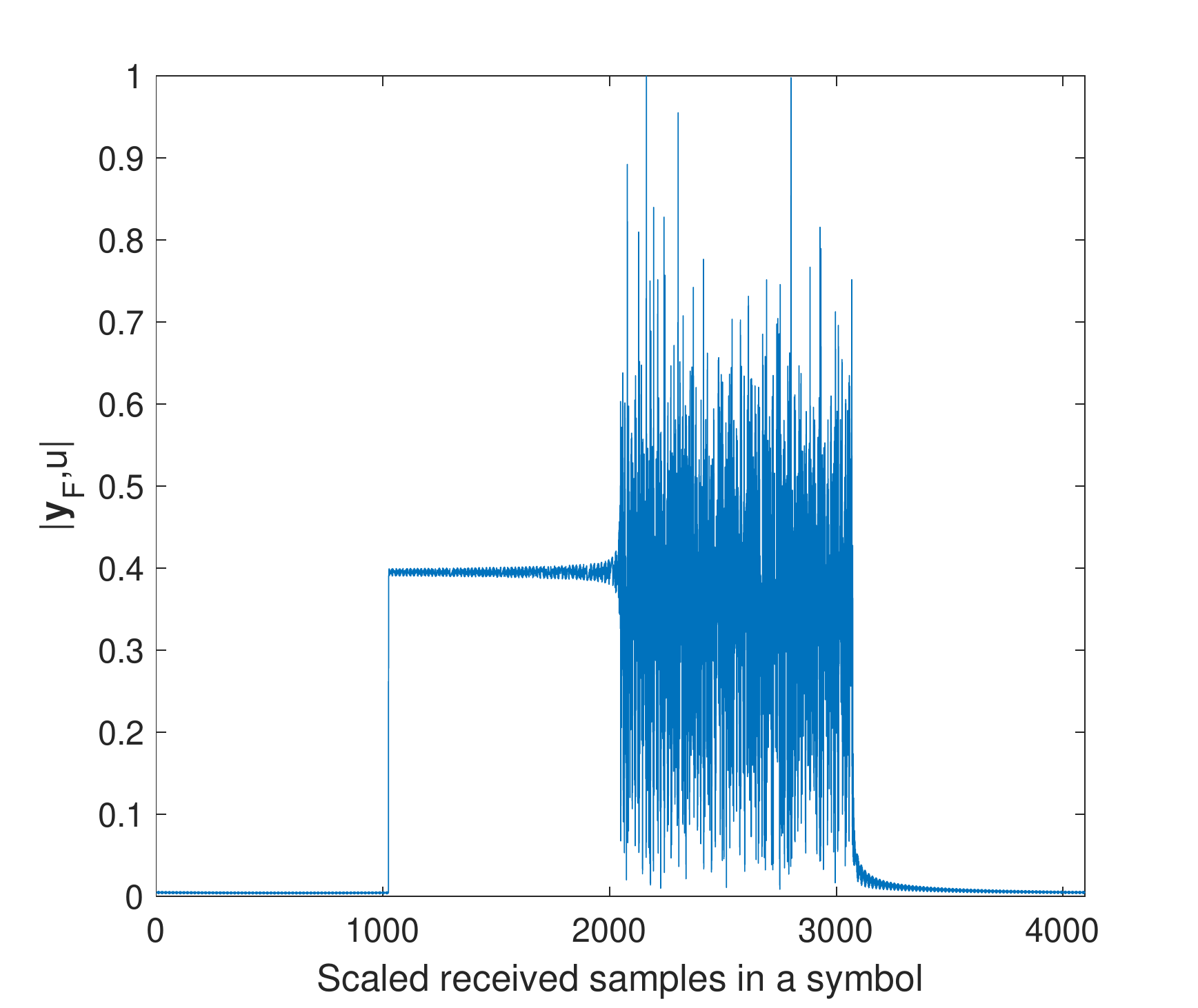}}
               \subfloat[30 kHz numerology]{\label{fftwithout15khz:2}\includegraphics[width=45mm]{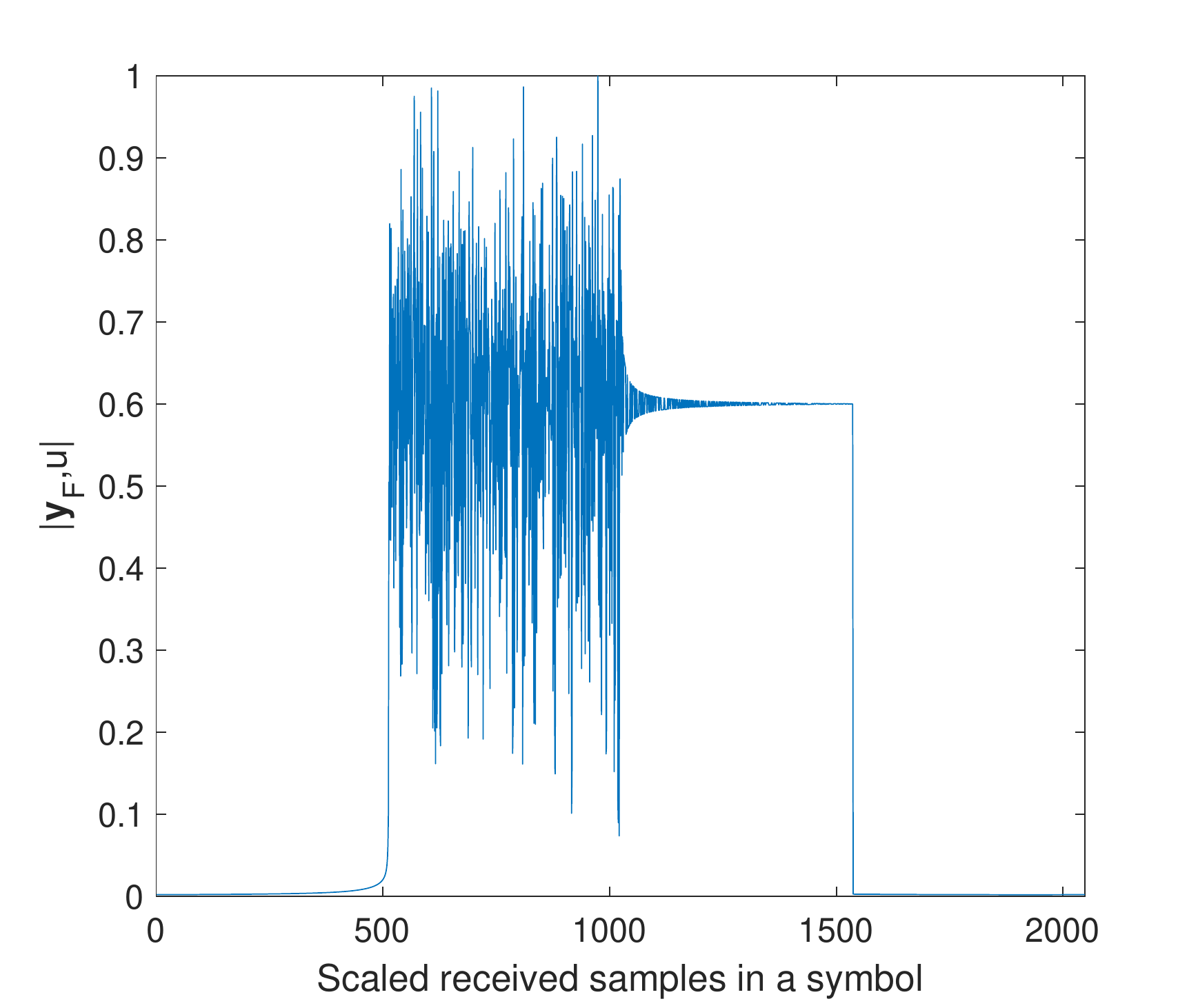}}  
                \\
                \end{center}
                 \centering
           \label{fftwithout15khz}\caption{Amplitude variations plots for Scenario 1 without considering noise and channel effects.}
\end{figure}

Algorithm 2 is also simulated for the Scenario 1 without taking channel distortion and noise into consideration, as shown in Fig. 3.
\thinspace It shows that the subband locations of the candidate numerologies can be easily identified and differentiate from each other.
{For example, the {calculated} $V_u$ values in Fig. \ref{fftwithout15khz:2} for the used band are 5.0733 and 0.0051, respectively. Therefore, it is straightforward to say that the subband location of the 30 kHz candidate numerology is the one with the lowest $V_u$ value (i.e. the second used band with $V_u=0.0051$).}

\begin{figure}[t]
\centering

\includegraphics[height=2.8in,width=3.6in]{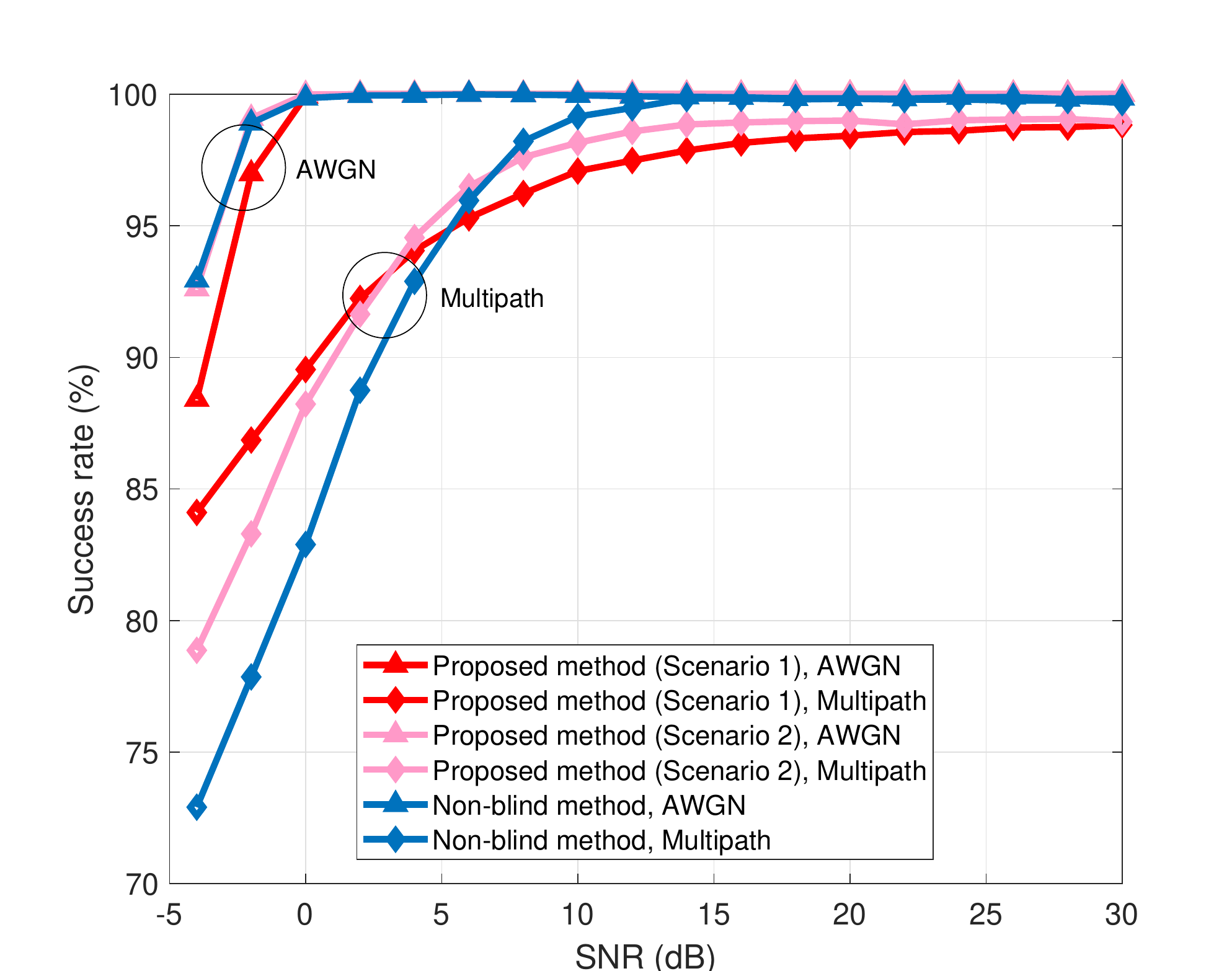}


\caption{The average success rate of detecting the candidate numerologies in the considered scenarios using the proposed algorithms and the non-blind identification approach under AWGN and Rayleigh fading channel.}

\label{accFDall}
\end{figure}

The performance of the proposed method is compared with {the non-blind approach in the traditional mixed-numerology system} under AWGN and multipath Rayleigh channel. {It is worth noting that prior information about the candidate numerologies is known in the non-blind identification approach.}
\thinspace Fig. \ref{accFDall} shows that the proposed algorithms provide robust accuracy results for a wide range of SNR {values}.
\thinspace More specifically, the proposed blind identification method and {the non-blind approach} ensure a success rate of 100\% for the considered scenarios at 0 dB of SNR under AWGN.
\thinspace {Fig. \ref{accFDall} also shows} that the types and locations of candidate numerologies can be detected with robust accuracy in multipath conditions.

\begin{figure}[t!]
\centering

\includegraphics[height=2.8in,width=3.6in]{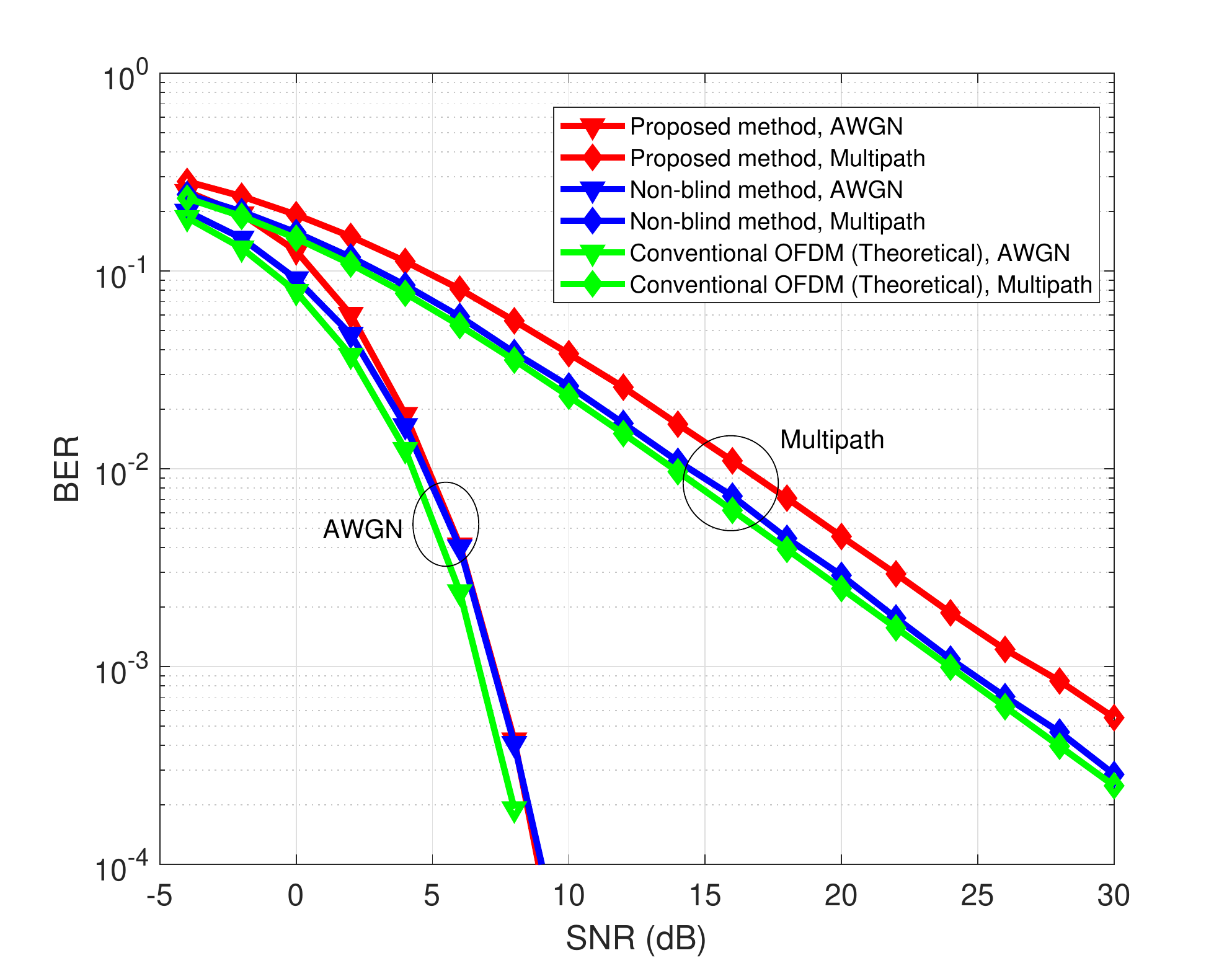}

\caption{Comparison of the average BER performance of the blind/non-blind approaches in the conventional mixed-numerology system, to the theoretical BER results of the conventional OFDM under AWGN and Rayleigh fading channel.}

\label{berallsch}
\end{figure} 

The BER performance of the proposed {blind identification} method is compared with {the non-blind identification approach in the} conventional mixed-numerology OFDM system. 
It should be noted that Scenario 1 is considered in the BER performance evaluation for the proposed method since it has worse success rate compared to the second one. As a benchmark, we also include the theoretical BER performances of the conventional OFDM {\cite{proakis2008digital} where single numerology is used}.
Fig. \ref{berallsch} shows that the proposed method is slightly worse than the theoretical BER {result} of the conventional OFDM under AWGN environment. Also, a slight BER difference between conventional OFDM and the {non-blind identification approach in the conventional mixed-numerology system} can be observed due to inserting sufficient guard bands to eliminate the effects of inter-numerology interference (INI), especially at high SNR values. This interference can be also decreased {by other} approaches \cite{8861343}, but this is out of the scope of this work. The additional blind operations performed in the proposed method introduce indispensable bits in errors. Thus, there is approximately $3$ dB loss in BER performance {of the proposed {blind} method} compared to {that of} the {non-blind one}.

\section{Conclusion} 
A new identification {method} for mixed-numerology OFDM is proposed based on the  characteristics of mixed-numerology signals in time and frequency domains. The success rates for numerology and subband location identification are {very high over} AWGN and frequency selective channels. By using the proposed method, the numerology and the subband location can be detected without prior information, {but with} a slight BER performance loss.  
As future work, further accuracy improvements will be investigated, especially for frequency-selective channels and asynchronous transmission. 
\section*{Acknowledgment} The authors thank Salah Eddine Zegrar, Research Assistant, for assistance and comments that improved the manuscript.


\end{document}